\documentclass[12pt,preprint]{aastex}
\def\farcs{\hbox{$.\!\!^{\prime\prime}$}}
\def\farc{\hbox{$\ \!\!^{\prime\prime}$}}
\def\farcm{\hbox{$\!\!^{\ \prime}$}}

\slugcomment{}
\shorttitle{molecular gas in NGC~4945}
\shortauthors{Richard et al.}
\begin{document}


\title{The Circumnuclear molecular gas in the Seyfert Galaxy NGC~4945}


\author{Richard C. Y. Chou\altaffilmark{1,2} }
\affil{Institute of Astronomy and Astrophysics, Academia Sinica, P.O. Box 23-141, Taipei}
\affil{Department of Astronomy and Astrophysics, University of Toronto, 50 St. Geroge Street, Toronto, ON M5S 3H8}
\email{chou@astro.utoronto.ca}

\author{A. B. Peck\altaffilmark{3}}
\affil{Harvard-Smithsonian Center for Astrophysics, 60 Garden Street, Cambridge, MA 02138}
\email{apeck@cfa.harvard.edu}

\author{J. Lim\altaffilmark{1}}
\affil{Institute of Astronomy and Astrophysics, Academia Sinica, P.O. Box 23-141, Taipei}
\email{jlim@asiaa.sinica.edu.tw}

\author{S. Matsushita\altaffilmark{1}}
\affil{Institute of Astronomy and Astrophysics, Academia Sinica, P.O. Box 23-141, Taipei}
\email{satoki@asiaa.sinica.edu.tw}

\author{S. Muller\altaffilmark{1}}
\affil{Institute of Astronomy and Astrophysics, Academia Sinica, P.O. Box 23-141, Taipei}
\email{muller@asiaa.sinica.edu.tw}

\author{S. Sawada-Satoh\altaffilmark{1}}
\affil{Institute of Astronomy and Astrophysics, Academia Sinica, P.O. Box 23-141, Taipei}
\email{satoko@asiaa.sinica.edu.tw}

\author{Dinh-V-Trung\altaffilmark{1}}
\affil{Institute of Astronomy and Astrophysics, Academia Sinica, P.O. Box 23-141, Taipei}
\email{trung@asiaa.sinica.edu.tw}

\author{F. Boone\altaffilmark{4}}
\affil{Observatoire de Paris, LERMA, 61, av. de l'Observatoire, F75014 Paris}
\email{Frederic.Boone@obspm.fr}

\and

\author{C. Henkel\altaffilmark{5}}
\affil{ Max-Planck-Institut f\"{u}r Radioastronomie, Auf dem H\"{u}gel 69, 53121 Bonn, Germany}
\email{p220hen@mpifr-bonn.mpg.de}




\begin{abstract}
We have mapped the central region of NGC~4945 in the $J=2\rightarrow1$ transition of $^{12}$CO, $^{13}$CO, and C$^{18}$O, as well as the continuum at 1.3~mm, at an angular resolution of $5\farc \times 3\farc$ with the Submillimeter Array.  The relative proximity of NGC~4945 (distance of only 3.8~Mpc) permits a detailed study of the circumnuclear molecular gas and dust in a galaxy exhibiting both an AGN (classified as a Seyfert~2) and a circumnuclear starburst in an inclined ring with radius $\sim$2\farcs5 ($\sim$50~pc).  We infer the systemic velocity $\sim$$585 {\rm \ km \ s^{-1}}$ from channel maps and PV-diagrams.  We find that all three molecular lines trace an inclined rotating disk with major axis aligned with that of the starburst ring and large-scale galactic disk, and which exhibits solid-body rotation within a radius of $\sim$5\farc\ ($\sim$95~pc).  The rotation curve flattens beyond this radius, and the isovelocity contours exhibit an S-shaped asymmetry suggestive of a highly inclined bar as has been invoked to produce a similar asymmetry observed on larger scales.  We infer an inclination for the nuclear disk of $62^{\circ} \pm 2^{\circ}$, somewhat smaller than the inclination of the large-scale galactic disk of $\sim$$78^{\circ}$.  The continuum emission at 1.3~mm also extends beyond the starburst ring, and is dominated by thermal emission from dust.  If it traces the same dust emitting in the far-infrared, then the bulk of this dust must be heated by star-formation activity rather than the AGN.  We discover a kinematically-decoupled component at the center of the disk with a radius smaller than $1\farcs4$ (27~pc), but which spans approximately the same range of velocities as the surrounding disk.  This component has a higher density than its surroundings, and is a promising candidate for the circumnuclear molecular torus invoked by AGN unification models.
\end{abstract}


\section{INTRODUCTION}
\label{intro}
NGC~4945 is a nearby, almost edge-on ($i = 78^{\circ}$), disk galaxy (type SB(s)cd or SAB(s)cd) that is a member of the Centaurus Group of galaxies \citep{wes79}.  Its distance, determined most accurately from the luminosity of stars at the tip of the red giant branch, is 3.82$\pm$0.31~Mpc \citep{kar06}.  This value is consistent with previous more cruder estimates of the distance to NGC~4945 (see the discussion in Bergman et al. 1992), and so henceforth we shall assume that 1\farc\ corresponds to 19~pc in NGC~4945.

The central region of NGC~4945 contains an active galactic nucleus (AGN) revealed most unambiguously by its strong and variable hard X-ray emission \citep{iws93,don96,mad00}.  H$_2$O megamaser emission has been detected from the nucleus, likely distributed in a disk with radius of $\sim$15~mas ($\sim$0.3~pc) around the AGN (Greenhill et al. 1997).  The AGN is classified as a Seyfert 2 \citep{braa97,mad00,sch02}, consistent with findings that Seyfert~2 but not Seyfert~1 galaxies exhibit water megamasers \citep[e.g., review by][]{lok05}.  Surrounding the AGN, there is an inclined circumnuclear starburst ring with a radius of $\sim$2\farcs5 ($\sim$50 pc) seen most clearly in Pa$\alpha$ \citep{mar00}.  The central region of NGC~4945 is among the strongest and most prolific extragalactic sources of molecular lines \citep[e.g.,][]{hen94,cur01,wan04}.  For these reasons, NGC~4945 is a particularly attractive candidate for studying the nature of molecular gas at the center of an active galaxy, and the role this gas plays in fueling and perhaps also determining in part the observed properties of both the circumnuclear starburst and AGN.

Because of its southerly declination ($\delta = -49^{\circ}$), studies of the molecular gas in NGC~4945 have been largely restricted to single-dish telescopes, in particular the 15-m Swedish-ESO Submillimeter Telescope (SEST).   \citet{whi90} made the first map of the central region of this galaxy in the $J=1\rightarrow0$ transition of $^{12}$CO at angular resolution of 43\farc\ (measured at full-width to half-maximum, FWHM, henceforth used throughout when quoting sizes).  They attributed the observed strong central concentration of molecular gas to a disk with a radius of $\sim$18\farc\ (340~pc).  With just a single pointing at the nucleus, \citet{ber92} obtained high-quality spectra of both $^{12}$CO and $^{13}$CO in the $J=1\rightarrow0$ transition.  They found that the observed line ratio as a function of velocity can be modeled as a homogenous ring in rigid-body rotation with an inner radius of $\sim$8\farc\ ($\sim$150~pc) and an outer radius of $\sim$15\farc\ ($\sim$280~pc).

\citet{dah93} mapped the central region of NGC~4945 in both the $J=1\rightarrow0$ and $J=2\rightarrow1$ transitions of $^{12}$CO at an angular resolution of 43\farc\ and 24\farc\ respectively.  Based on the measured position-velocity diagram along the major axis of NGC~4945 in both transitions, they inferred the presence of a ring with a radius of 8\farc$\pm$3\farc\ ($\sim$150~pc) in the $J=2\rightarrow1$ and 15\farc$\pm$3\farc\ ($\sim$280~pc) in the $J=1\rightarrow0$ transitions.  They noted that this did not necessarily suggest the presence of two rings, but rather a radial variation in the excitation conditions and/or optical depth in a single ring with a given radial thickness.

\citet{mau96} mapped the central region of NGC~4945 in the $J=3\rightarrow2$ transition of $^{12}$CO at an angular resolution of 15\farc.  In this transition, they found an intrinsic size for the central concentration of radius $5\farcs2 \pm 1\farcs5$ ($\sim$100~pc) along its major axis.  Although the position-velocity diagram is once again consistent with a ring but with an even smaller radius than in lower $^{12}$CO transitions, \citet{mau96} were careful to point out that other structures (e.g., disk, bar, spiral, or even two independent molecular concentrations) are possible.

All the abovementioned single-dish observations infer the presence of a molecular ring with size smaller than the respective angular resolutions of these observations.  It is not clear how this molecular ring is related to the starburst ring, whether it comprises the reservoir for fueling the active nucleus, and whether it is related to the hypothetical circumnuclear molecular torus required by AGN unification models \citep{ant93}.  Observations at higher angular resolutions are needed to spatially resolve and hence clarify the spatial-kinematic distribution of the molecular gas at the center of NGC~4945.  The first such observations were made quite recently by \citet{cun05}, who used the Australia Telescope Compact Array (ATCA) to map a central region of radius 17\farc\ (323~pc) in the $J=1\rightarrow0$ transitions of HCN, HCO$^+$, and HNC at an angular resolution of $5\farcs6 \times 3\farcs5$ ($106 \times 66$~pc).  These lines trace molecular gas at about an order of magnitude higher density ($\sim$$10^{4-5} {\rm \ cm^{-3}}$) than the abovementioned $^{12}$CO lines ($\sim$$10^{3-4} {\rm \ cm^{-3}}$).  The HNC map, which best traces the overall spatial-kinematic distribution of the molecular gas, reveals an inclined rotating disk-like feature with a radius of $4\farcs25 \pm 0\farcs25$ ($\sim$80~pc) along the major axis and position angle of $64^{\circ}$, somewhat misaligned from the major axis of both the starburst ring and large-scale galactic disk at a position angle of $\sim$$45^{\circ}$.  Based on the measured position-velocity diagram, \citet{cun05} argue that the HNC line traces a ring with an inner radius of 3\farc\ (57~pc).  Their maps, however, are affected by a lack of short baselines that result in an inability to detect structures larger than $\sim$15\farc\ ($\sim$285~pc). Furthermore, line absorption against the strong central continuum source at their observing wavelength of 3.3~mm (90~GHz) may compromise the detectability of the innermost features.

Here, we present the first interferometric observations of the central region of NGC~4945 in $^{12}$CO, as well as in $^{13}$CO and C$^{18}$O, at their $J= 2\rightarrow1$ transitions.  The angular resolution attained is much higher than previous single-dish CO observations, and we properly resolve for the first time the central molecular gas concentration in CO lines.  Our maps recover virtually all the emission detected in single-dish observations, and are not affected by line absorption against the central continuum source.  We show that the CO gas is distributed in an inclined disk exhibiting rigid-body rotation out to a radius of $\sim$5\farc\ ($\sim$95~pc); i.e., the molecular disk extends beyond the circumnuclear starburst ring seen in Pa$\alpha$.  The rotation curve flattens beyond this radius, and the isovelocity contours exhibit an S-shaped asymmetry suggestive of a highly inclined bar as has been invoked to produce a similar asymmetry in both molecular and atomic hydrogen gas extending to much larger radii.  We do not detect any central hole in the disk, placing an upper limit of 2\farcs2 (42~pc) on the radius of any such hole.  Instead, we detect for the first time a spatially-unresolved component at the center of the disk that is kinematically decoupled from the surrounding disk.  This inner component exhibits a broad velocity width comparable with the overall range in velocities exhibited by the surrounding disk, and is a good candidate for the hypothesized circumnuclear molecular torus invoked by AGN unification models \citep{ant93}.


\section{OBSERVATIONS AND DATA REDUCTION}
\label{obs}
We observed NGC~4945 with the Submillimeter Array \citep[SMA;][]{ho04}, located on Mauna Kea in Hawaii, on 2004 May 22 and 24.  To achieve the most circular beam possible, we used the compact configuration of the array that is elongated in the north-south direction tailored for southern hemisphere sources.  Over the course of the observations, this array configuration provided projected baselines ranging from about 6~m to 74~m.  Seven of the eight antennas of the array were in operation at the time of our observations.  Each of the SMA antennas has a diameter of 6~m, and at our observing wavelength of 1.3~mm (frequency of 230~GHz) a corresponding FWHM primary beam of 52\farc\ (i.e., field of view of nearly 1~kpc). 

The array was pointed toward the center of NGC~4945 as defined by the position of H$_2$O megamasers at right ascension $\alpha({\rm 2000}) = {\rm 13^{h} 05^{m} 27\fs5}$ and declination $\delta({\rm 2000}) = {\rm 49^{\circ} 28\farcm\ 05\farcs56}$ \citep{gre97}.  We observed the $^{12}$CO, $^{13}$CO, and C$^{18}$O lines in their $J=2\rightarrow1$ transitions simultaneously by placing the $^{12}$CO line in the upper sideband (USB) and the $^{13}$CO and C$^{18}$O lines in the lower sideband (LSB) of the double-sideband receivers.  The correlator was configured to provide a total bandwidth of 2~GHz in each sideband using twenty-four independent spectral windows (``chunks'').  Each window is divided into 128 channels with a channel width of 0.8125 MHz (i.e., $\sim$1.1 km/s), giving a total frequency coverage in each sideband (with a small overlap between windows at their edges) of 1.9 GHz (i.e., $\sim$2570~km/s).

Because of its low declination, NGC~4945 is visible for only $\sim$2~hrs each day from the SMA, and never exceeds an elevation of $\sim$$25^{\circ}$.  Although the weather was good (zenith opacity $\tau = 0.15 - 0.20$) on both days of our observations, because of its low elevation the sky opacity towards the source was relatively high resulting in system temperatures of 400~K to 600~K (single sideband).  On each day, we observed NGC~4945 for a total on-source integration time of 53~mins.  We used two quasars for complex gain (i.e., amplitude and phase) calibration.  The stronger quasar $1334-127$, which lies at a larger angular separation of $\sim$$40^{\circ}$ from NGC~4945, was used for amplitude calibration.  The weaker quasar $1313-333$, which lies at a smaller angular separation of $\sim$$20^{\circ}$ from NGC~4945, was used for phase calibration.  The flux density of the amplitude calibrator $1334-127$ was scaled from observations of Callisto.  The quasar 3C279 was observed for bandpass calibration.

We calibrated the data in the standard fashion using the software package MIR-IDL, and made images using AIPS.  The results shown in this paper correspond to the data taken on May 24 only.  Data taken on May 22, at a different sidereal time, have a different $uv$-coverage (greater number of longer projected baselines) and deep negative sidelobes that made imaging of the line emission difficult.  To make a map of the continuum emission, we used the line-free channels (total bandwidth of 3~GHz) in both the LSB and USB.  To make maps in line emission, we first used the line-free channels in the visibility plane to derive the continuum emission, which was then subtracted from all the channels.  The synthesized beam achieved with natural weighting of the visibilities is $5\farcs1 \times 2\farcs8$ ($\sim$$100 \times 50$~pc) at a position angle of $8\fdg8$.

\section{RESULTS AND INTERPRETATION}\label{sec:results}
\subsection{Continuum Emission}\label{subsec:continuum} 

\subsubsection{Contribution from dust}
\begin{figure}[htb]
\begin{center}
\includegraphics[width=12cm,angle=0]{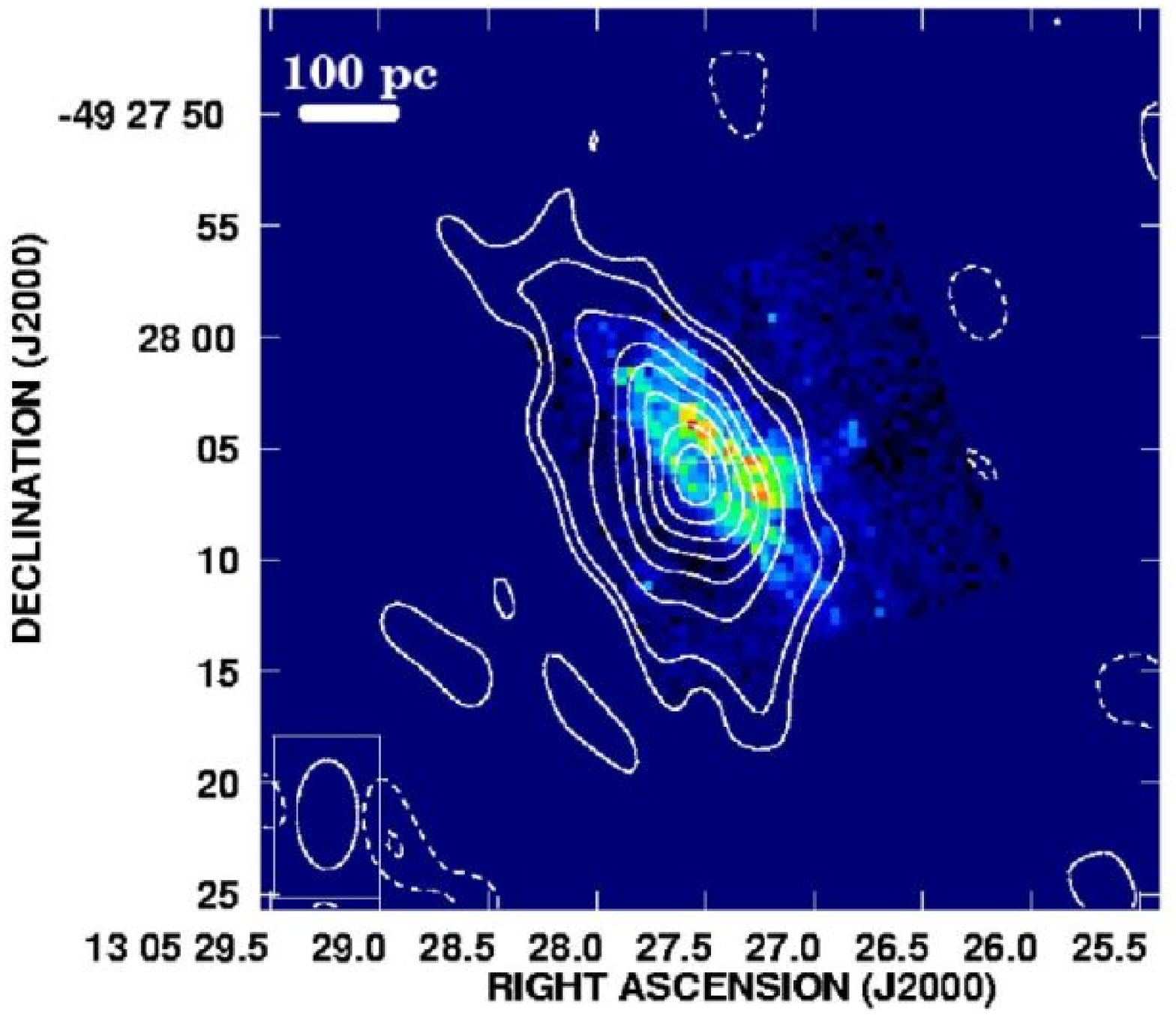}
\end{center}
\caption\small{Contours of the continuum emission at 1.3~mm measured with the SMA superposed on a negative Pa$\alpha$ image of the central region of NGC~4945 taken from Marconi et al. (2000).  Contour levels are plotted at $-3\sigma$, $-2\sigma$, $2\sigma$, $3\sigma$, and then in steps of $3\sigma$ to $21\sigma$, where $\sigma$ is the root-mean-square (rms) noise level of $13$ mJy/beam.  The main elongated feature in the Pa$\alpha$ image is the starburst ring.  The cross denotes the position of the AGN as determined from the position of H$_2$O megamasers \citep{gre97}.  The synthesized beam is indicated at the lower left corner and has a size of $5\farcs1 \times 2\farcs8$ and a position angle of $8^{\circ}.8$.}
\label{continuum}
\end{figure}

In Figure~\ref{continuum}, we show a map of the 1.3-mm continuum emission superposed on a Pa$\alpha$ image of the central region of NGC~4945 taken from \citet{mar00}.  The continuum source has a total flux density of $1.3 \pm 0.2$~Jy, and is clearly resolved.  Gaussian fitting yields a deconvolved size (at FWHM) of $9\farcs8 \times 5\farcs0$ $(\pm 0\farcs3)$ ($186 \times 95$~pc) with the major axis at a position angle (PA) of $28^{\circ} \pm 3^{\circ}$.  The centroid of this source is located at $\alpha({\rm 2000}) = {\rm 13^{h} 05^{m} 27\fs59 \pm 0\fs01}$ and $\delta({\rm 2000}) = {\rm -49^{\circ} 28\farcm\ 06\farcs1 \pm 0\farcs2}$, which is 1\farcs8 to the south-east of the nominal centroid of the H$_2$O megamasers \citep{gre97}.  The latter coincides within measurement uncertainties with the centroid of the central radio continuum source detected at cm-wavelengths \citep{elm97}, and presumably marks the location of the AGN.  

The major axis of the 1.3-mm continuum source is not aligned with the major axis of either the starburst ring or larger-scale galactic disk, both of which have a ${\rm PA} \approx 45^{\circ}$.  Along the major axis of the Pa$\alpha$ starburst ring, the 1.3-mm continuum extends beyond the inner bright rim of the starburst ring at a radius of $\sim$2\farcs5 ($\sim$50~pc), and also beyond the detectable outer radius of this ring at $\sim$5\farc\ ($\sim$100~pc). On the south-eastern side of the ring, the 1.3-mm continuum clearly extends beyond the measured extent of the Pa$\alpha$ emission.

The central region of NGC~4945 has been imaged in the continuum at centimeter wavelengths at angular resolutions comparable with that attained here.  At 21~cm (1.4~GHz), the source has a deconvolved size of $7\farcs6 \times 3\farcs4$ $(\pm 0\farcs2)$ with major axis at ${\rm PA} = 42^{\circ}.5$, and a total flux density of $4.6 \pm 0.1$~Jy \citep{ott01}.  At 5~cm (6~GHz), the source has a smaller deconvolved size of $5\farcs7 \times 2\farcs0$ ($\pm 0\farcs1$) at ${\rm PA} = 43^{\circ} \pm 1^{\circ}$, and a total flux density of $2.04 \pm 0.04$~Jy \citep{ww90}.  The size of the continuum source at centimeter wavelengths is therefore much smaller than that measured at 1.3~mm.  Instead, the centimeter continuum source has a size comparable with the starburst ring, and its major axis is aligned with that of the starburst ring.  The steep negative spectral index of this source indicates that nonthermal (synchrotron) emission dominates at centimeter wavelengths, and presumably arises from star-formation-related activity (e.g., supernovae) in the starburst ring (in addition to any unresolved emission from the AGN).

\citet{cun05} have imaged the central continuum source at 3.3~mm (90~GHz) at an angular resolution comparable with that attained here.  Like us, they find an elongated source whose centroid is offset to the south-east of the H$_2$O megamasers, and whose major axis is at ${\rm PA} \approx 29^{\circ}$.  With a reported deconvolved source size of $7\farcs6 \times 2\farcs0$ and peak flux density of 0.13~Jy, the corresponding total flux density for a Gaussian source is $\sim$1.0~Jy.  Extrapolating from 21~cm and 5~cm, the expected flux density at 3.5~mm is $\sim$0.5~Jy, only half that actually measured.  Extrapolating to 1.3~mm, the estimated contribution from nonthermal emission is $\sim$0.3~Jy, less than one quarter the flux density measured at this wavelength.  Even if we assume that nonthermal emission dominates at 3.3~mm, this emission would only contribute at most about half the total flux density measured at 1.3~mm.  Thus, we conclude that dust emission dominates at 1.3~mm (and also contributes significantly at 3.3~mm), and that this explains the different dimensions and position angles of the source at millimeter and centimeter wavelengths.  In $\S\ref{sec:far-ir}$, we discuss the implications of these results for the origin of the central far-IR emission from NGC~4945.

\subsubsection{Molecular gas mass from dust}
\label{subsec:dustmass}
To estimate the mass of molecular gas from the inferred dust emission at 1.3~mm (assuming a gas to dust ratio of 100), we first subtract the estimated cotribution from nonthermal emission ($\sim$0.3~Jy) from the total continuum emission ($\sim$1.3~Jy).  The gas mass is then given by
\\
\begin{equation}
M_{gas} (M_\odot) = 1730 \frac{S_\nu[{\rm mJy}] D[{\rm Mpc}]^2 \lambda[{\rm mm}]^2} {T_d[{\rm K}] \kappa_d(\nu)[{\rm \ g^{-1} \ cm^2}]}
\end{equation}
\\
\noindent \citep{hil83}, where the dust continuum flux density $S_\nu \approx 1 {\rm \ Jy}$, distance $\rm D = 3.82~Mpc$, wavelength $\rm \lambda=1.3~mm$, and dust absorption coefficient $\kappa_d(1.3 \rm{mm}) \approx 3 \times 10^{-3} {\rm \ g^{-1} \ cm^2}$.  Assuming a dust temperature $T_d \approx 40 {\rm \ K}$ as inferred from far-infrared measurements \citep{bro88}, we find that $M_{\rm gas} \approx 3.6 \times 10^8$~M$_\odot$.  If we adopt a dust temperature corresponding to the peak brightness temperature measured in $^{12}$CO(2-1) of $\sim$30~K (\S\ref{subsec:lineemission}), the gas mass is then $M_{\rm gas} \approx 4.7 \times 10^8$~M$_\odot$.  As we shall show in $\S$\ref{co-mass}, the gas mass inferred from dust is comparable with values inferred from the three observed CO lines.

\subsection{Line emission}\label{subsec:lineemission} 

\subsubsection{Spatial-kinematic structure}
\label{spatial-kinetic-structure}
\begin{figure}[htb]
\begin{center}
\includegraphics[width=17cm,angle=0]{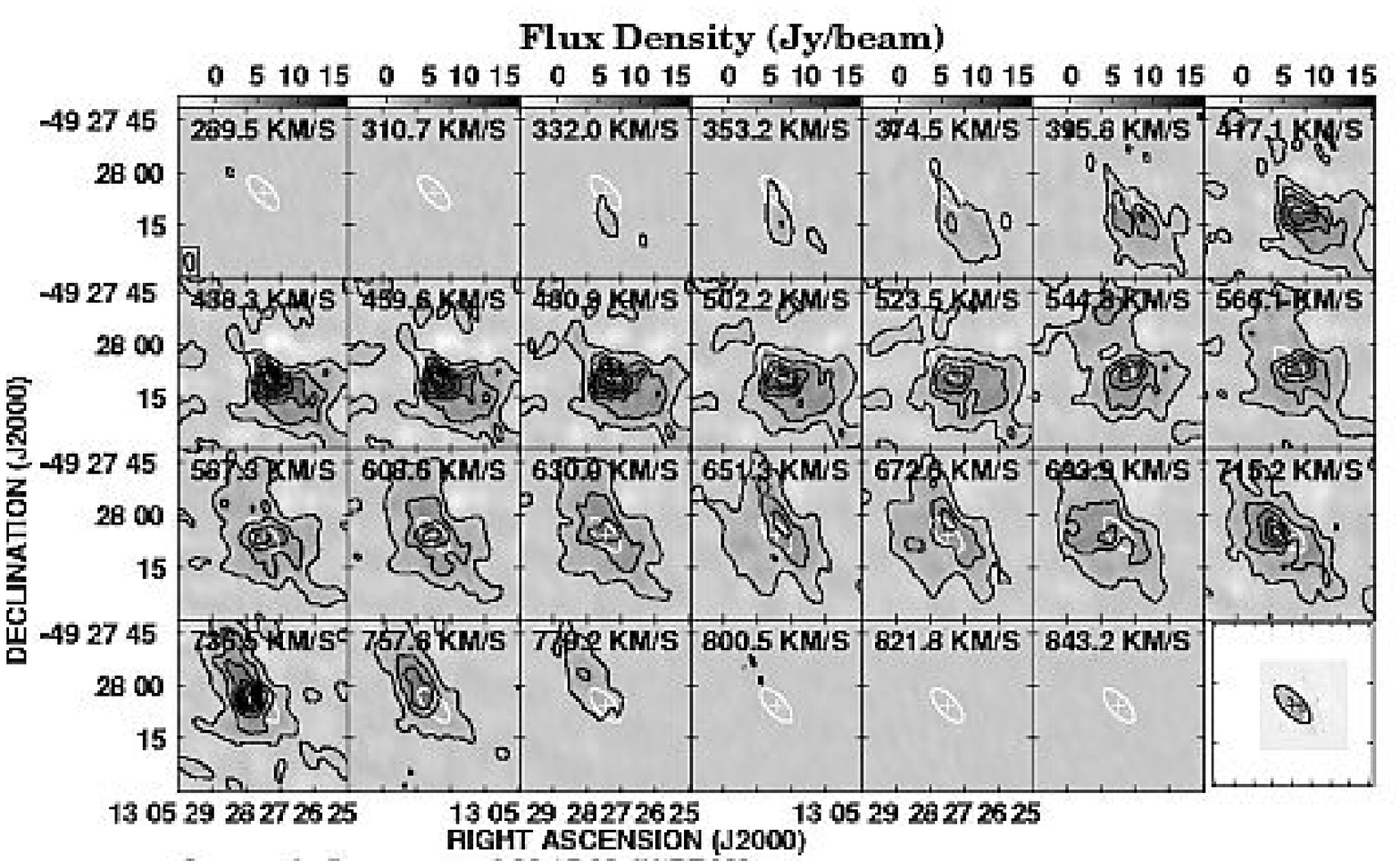}
\end{center}
\caption\small{$^{12}$CO(2-1) channel maps with contours plotted from $3\sigma$ to $75\sigma$ in steps of 12$\sigma$, where $\sigma = 0.2$~Jy/beam.  The velocity indicated in each panel is measured relative to the local standard of rest, which is $4.6 {\rm \ km \ s^{-1}}$ lower than the heliocentric velocity.  The velocity separation between each panel is $\sim$$20 {\rm \ km \ s^{-1}}$.  The cross denotes the location of the AGN, and the ellipse marks the approximate size of the starburst ring as shown in the bottom right panel.  The synthesized beam is shown at the lower left corner of the top left panel, and has a size of $5\farcs1 \times 2\farcs8$ and a position angle of $8^{\circ}.8$.}
\label{12channel}
\end{figure}

\begin{figure}[htb]
\begin{center}
\includegraphics[width=16cm,angle=0]{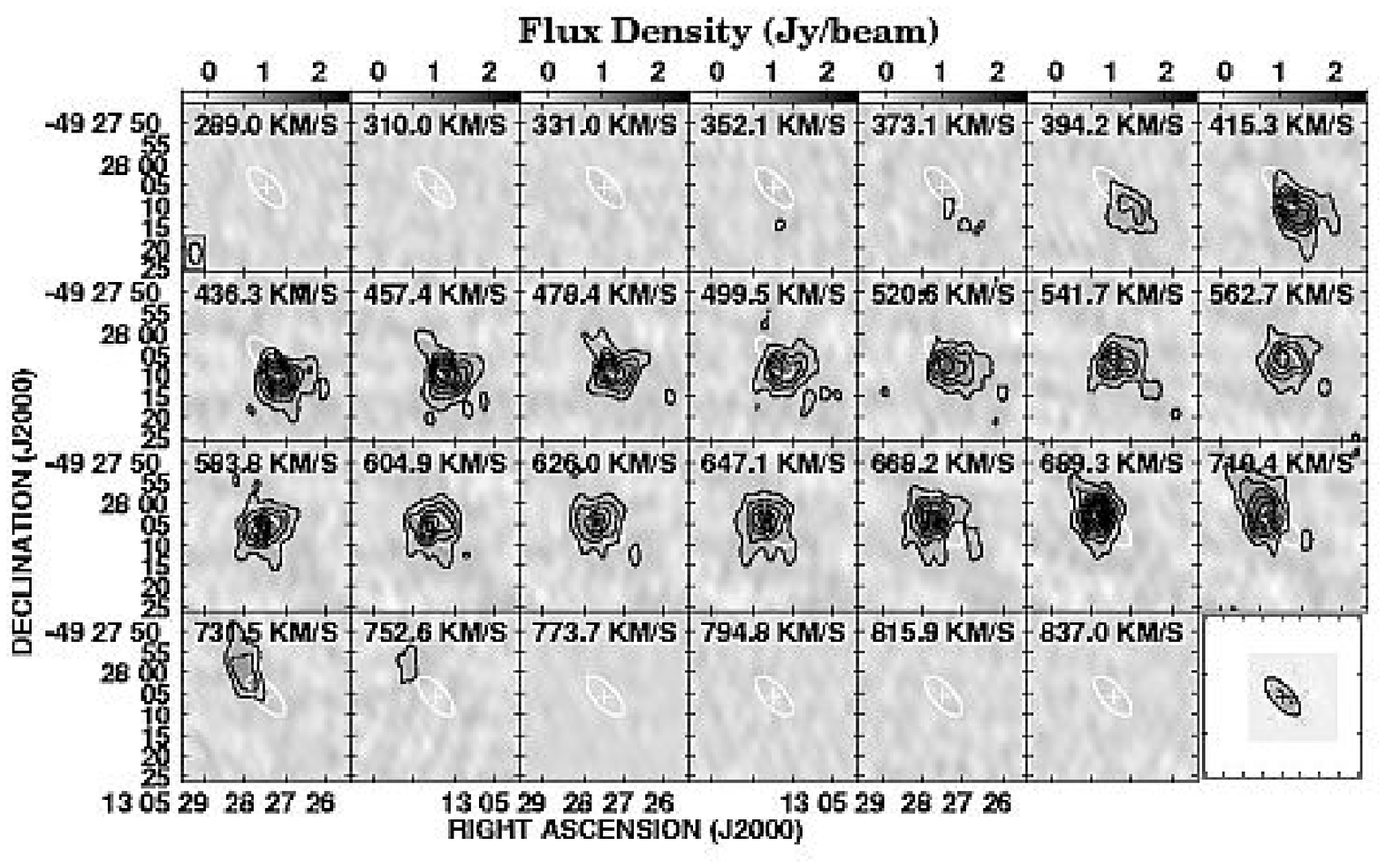}
\end{center}
\caption\small{$^{13}$CO(2-1) channel maps with contours plotted from $3\sigma$ to $24\sigma$ in steps of 3$\sigma$, where $\sigma = 0.1$~Jy/beam.  The synthesized beam is shown at the lower left corner of the top left panel, and has a size of $5\farcs0 \times 2\farcs8$ and a position angle of $0^{\circ}.7$. The rest are the same as Figure~\ref{12channel}.}
\label{13channel}
\end{figure}

\begin{figure}[htb]
\begin{center}
\includegraphics[width=16cm,angle=0]{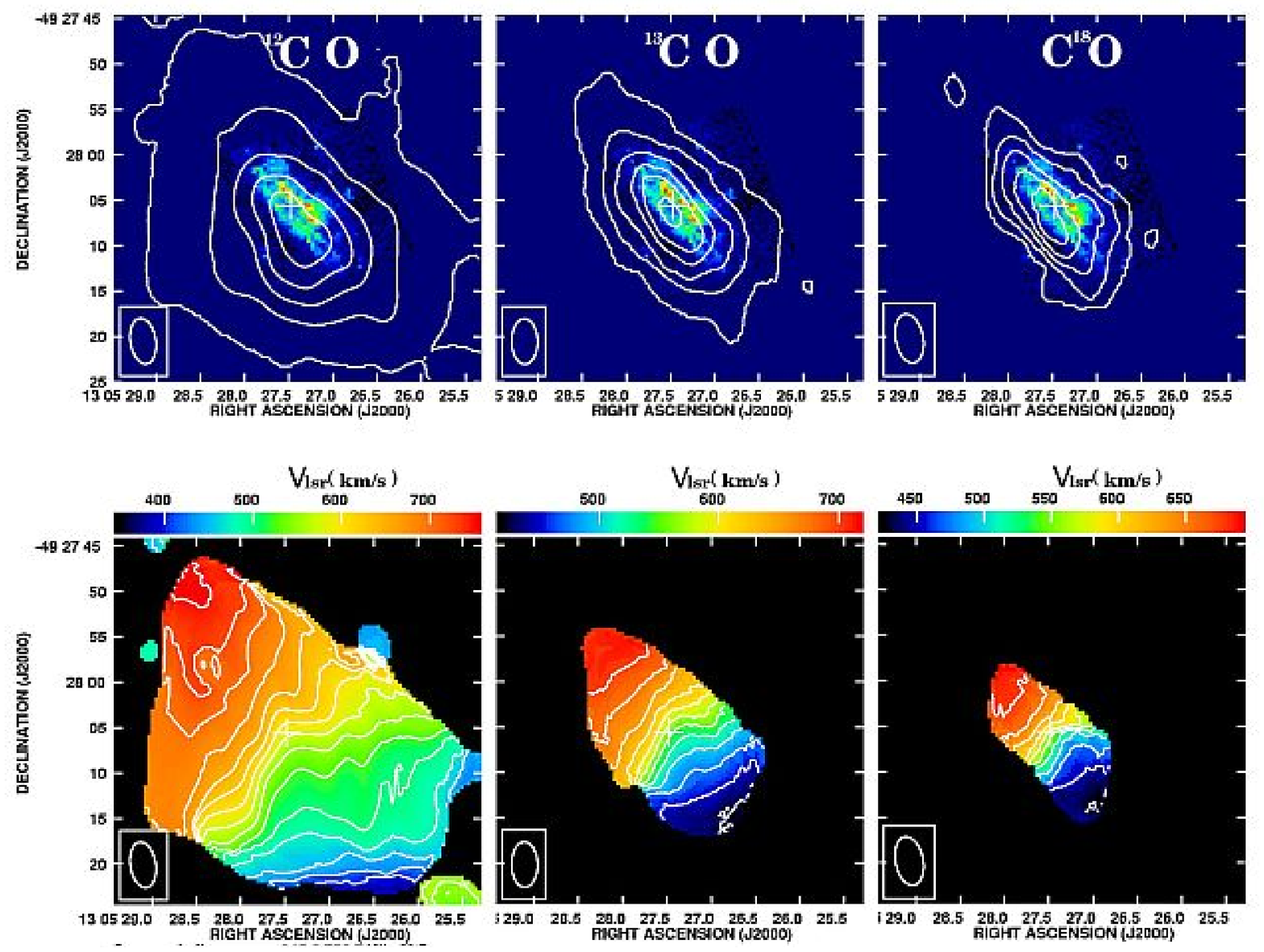}
\end{center}
\caption\small{Upper row, left to right: Contours of integrated intensity maps in $^{12}$CO(2-1), $^{13}$CO(2-1) and C$^{18}$O(2-1) overlaid on the Pa$\alpha$ image from Marconi et al. (2000).  Contour levels are plotted from 3$\sigma$ with steps of 320$\sigma$, 32$\sigma$ and 16$\sigma$ for $^{12}$CO(2-1), $^{13}$CO(2-1) and C$^{18}$O(2-1), respectively.  The ellipse at the lower left corner indicates the synthesized beam.  The cross denotes the position of the AGN.  Lower row, left to right: Corresponding intensity-weighted mean-velocity maps in $^{12}$CO(2-1), $^{13}$CO(2-1) and C$^{18}$O(2-1).  The velocity measured with respect to the local standard of rest is indicated by the horizontal bar at the upper edge of each panel.  The isovelocity contours are plotted from $400$~${\rm \ km \ s^{-1}}$~to $800$~${\rm \ km \ s^{-1}}$~in steps of $25$${\rm \ km \ s^{-1}}$.}
\label{mom0+1-maps}
\end{figure}

Like the continuum emission, the $^{12}$CO(2-1), $^{13}$CO(2-1), and C$^{18}$O(2-1) lines are strongly concentrated towards the center of NGC~4945.  Figures~\ref{12channel} and \ref{13channel} show channel maps of the $^{12}$CO(2-1) and $^{13}$CO(2-1) emission smoothed to a velocity resolution of $\sim$$20 {\rm \ km \ s^{-1}}$, with the circumnuclear starburst ring seen in Pa$\alpha$ plotted as an ellipse.  In Figure~\ref{mom0+1-maps}, we show the corresponding total intensity (moment 0) as well as intensity-weighted mean-velocity (moment 1) maps. 

An inspection of these maps reveals that the molecular emission originates (primarily) from a highly-inclined rotating disk extending beyond the Pa$\alpha$ starburst ring.  Gaussian fitting to the moment maps yields a deconvolved size (at FWHM) for the disk of $16\farcs4 \times 10\farcs8$ ($\pm 0\farcs1$) ($\sim$$310 \times 205$~pc) and major axis at PA$=38^{\circ} \pm 1^{\circ}$ in $^{12}$CO(2-1), a size of $14\farcs2 \times 6\farcs6$ ($\pm 0\farcs1$) ($\sim$$270 \times 125$~pc) and PA$=43^{\circ} \pm 1^{\circ}$ in $^{13}$CO(2-1), and a size of $13\farcs1 \times 4\farcs7$ ($\pm 0\farcs1$) ($\sim$$250 \times 90$~pc) and PA$=45^{\circ} \pm 1^{\circ}$ in C$^{18}$O(2-1).  The derived position angle of the major axis in the more optically thin $^{13}$CO(2-1) and C$^{18}$O(2-1) lines is in good agreement with the position angle along which the velocity gradient of the disk reaches a maximum in these lines.  The major axis of the disk (mean ${\rm PA}=44^{\circ}$) is therefore well aligned with that of the starburst ring as well as the larger-scale galactic disk at a position angle of $\sim$$45^{\circ}$.  We see no evidence for the central hole in our data, suggesting that any central hole in this disk has a size smaller than the synthesized beam, which measures 4\farcs4 along the major axis of the disk.

In all three molecular lines, the inner region of the disk within a radius of $\sim$5\farc\ ($\sim$95~pc) appears to exhibit simple circular rotation with the isovelocity contours perpendicular to major axis of the disk as can be seen in Figure~\ref{mom0+1-maps}.  Beyond this radius, however, the disk exhibits significant deviations from circular rotation as can be best seen in $^{12}$CO(2-1).  Here, the isovelocity contours on the north-eastern side of the disk twist to the north, and on the south-western side twist to the south, producing a S-shaped asymmetry.  In this outer region, the spatial structure of the disk closely resembles the distribution of the 1.3-mm continuum emission (e.g., the south-east extension), reinforcing our earlier argument that the continuum emission at 1.3~mm is likely dominated by dust.

\begin{figure}[htb]
\begin{center}
\includegraphics[width=13cm,angle=0]{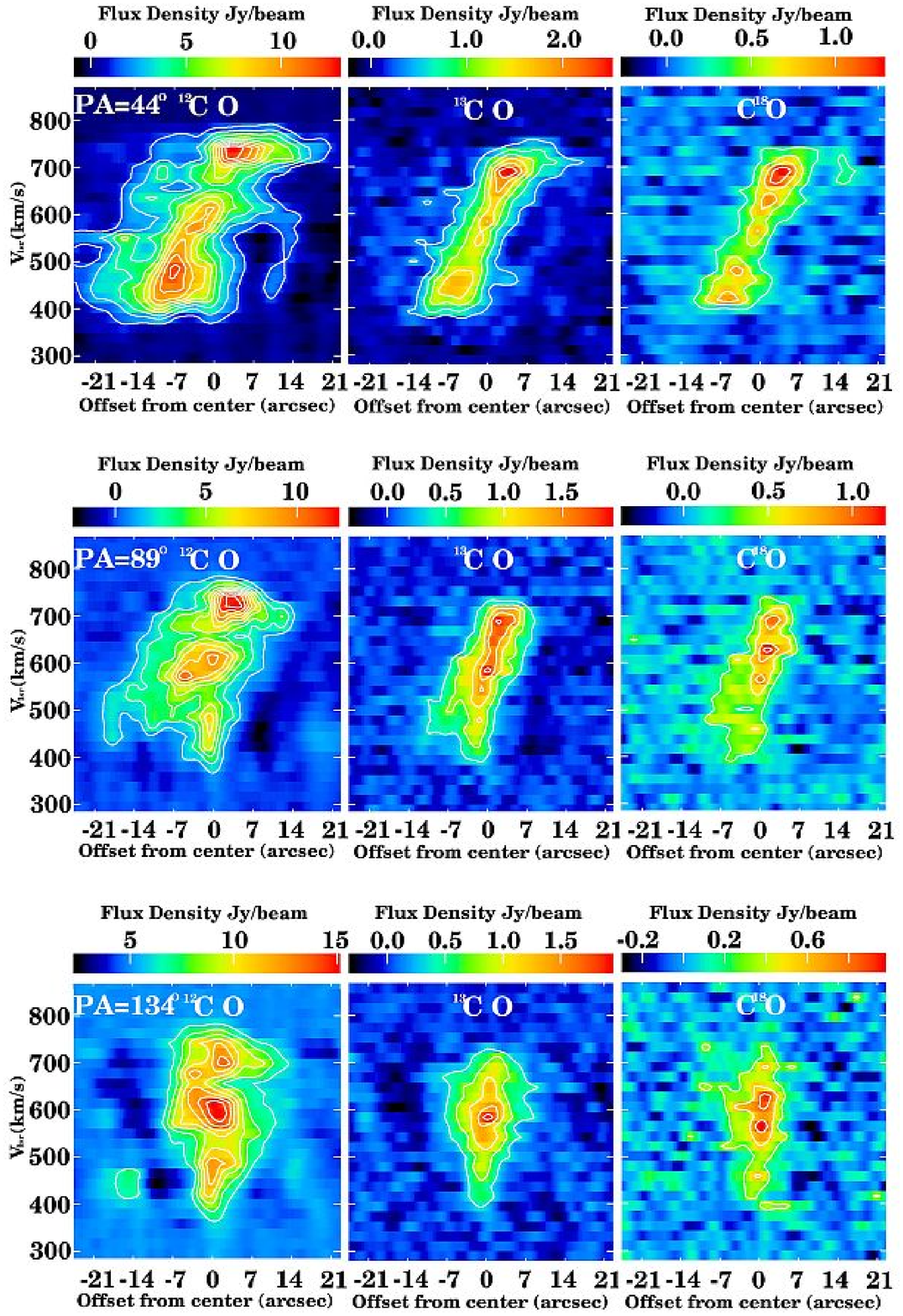}
\end{center}
\caption\small{Top row, left to right: Position-velocity diagram in $^{12}$CO(1-0), $^{13}$CO(2-1), and C$^{18}$O(2-1) along the major axis of the disk (${\rm PA}=44^{\circ}$).  Middle row, left to right: Same, but with a PA halfway between the major and minor axis of the disk (${\rm PA}=89^{\circ}$).  Bottom row, left to right: Same, but along the minor axis of the disk (${\rm PA}=134^{\circ}$).  In all panels, contour levels for $^{12}$CO are plotted from 7$\sigma$ in steps of 7$\sigma$, and contour levels for $^{13}$CO and C$^{18}$O from 3$\sigma$ in steps of 3$\sigma$.  All velocities quoted here are relative to the local standard of rest, which is $4.6 {\rm \ km \ s^{-1}}$ lower than the heliocentric velocity.}
\label{pv-diagrams}
\end{figure}

The position-velocity (PV-) diagrams of all three molecular lines along the inferred major axis of the disk are shown in Figure~\ref{pv-diagrams}.  The PV-diagrams along the major axis in $^{13}$CO(2-1) and C$^{18}$O(2-1) indicate (primarily) rigid-body rotation, while in $^{12}$CO(2-1) it shows additional complex emission with a broad velocity width at or near disk center.  Cuts along different position angles reveal that this ``excess'' emission is always present as a (vertical) strip in velocity at the origin of the PV-diagram, not just in $^{12}$CO(2-1) but also in $^{13}$CO(2-1) and C$^{18}$O(2-1).  For example, in Figure~\ref{pv-diagrams} we also show the PV-diagrams along the minor axis of the disk ($\rm PA = 134^{\circ}$), as well as halfway between the major and minor axes of the disk ($\rm PA = 89^{\circ}$), in all three molecular lines.  Armed with this knowledge, a closer inspection reveals that this feature can just be seen in the PV-diagrams along the minor axis of the disk in $^{13}$CO(2-1) and C$^{18}$O(2-1).  This feature therefore corresponds to a kinematically-decoupled component with a size smaller than the synthesized beam (i.e., projected radius as small as 1\farcs4 or 27~pc) located at the center of the disk.  In the channel maps, this kinematically-decoupled feature appears as the ever-present emission towards the center of the disk even at the most blueshifted and redshifted velocities detectable.  By contrast, in an inclined disk exhibiting only rigid-body rotation, the most blueshifted and redshifted velocities should originate from just the outermost regions of the disk along its major axis.

We infer from both the channel maps and PV-diagrams a systemic velocity of $\sim$$585 {\rm \ km \ s^{-1}}$ measured with respect to the local standard of rest.  (All velocities quoted here are relative to the local standard of rest, which is $4.6 {\rm \ km \ s^{-1}}$ lower than the heliocentric velocity.)  By contrast, single-dish observations in $^{12}$CO(1-0) and $^{12}$CO(2-1) infer a significantly lower systemic velocity of $\sim$$558 {\rm \ km \ s^{-1}}$ \citep{whi90,ber92,dah93}.  The HNC observation of \citet{cun05} indicate a systemic velocity, as measured at the midpoint between the two strongest HNC peaks in the PV-diagram (their Fig.~11), of $\sim$$570 {\rm \ km \ s^{-1}}$, about halfway between single-dish and our interferometric CO measurements.  The systemic velocity inferred from interferometric observations in atomic hydrogen (HI) gas \citep{ott01}, averaged over the entire galaxy, is $557 \pm 3 {\rm \ km \ s^{-1}}$, in agreement with the abovementioned value derived from single-dish CO measurements.  On the other hand, HI detected in absorption towards the centroid of the central continuum source has a velocity of $\sim$$585 {\rm \ km \ s^{-1}}$ \citep[see Fig.~5 of][]{ott01}, similar to the systemic velocity inferred here.  Arcsecond imaging of hydrogen recombination lines near 8.6~GHz \citep{roy05} shows that the emission peaks at the location of the central radio continuum source; these recombination lines have a systemic velocity of $\sim$$576 {\rm \ km \ s^{-1}}$, again close to that inferred here.  We therefore conclude, based on observations that better trace gas closer to the center of the galaxy, that the systemic velocity of NGC~4945 is more likely about $585 {\rm \ km \ s^{-1}}$.

\subsubsection{Comparison with single-dish observations}\label{subsec:singledish}
\begin{figure}[htb]
\begin{center}
\includegraphics[width=12cm,angle=0]{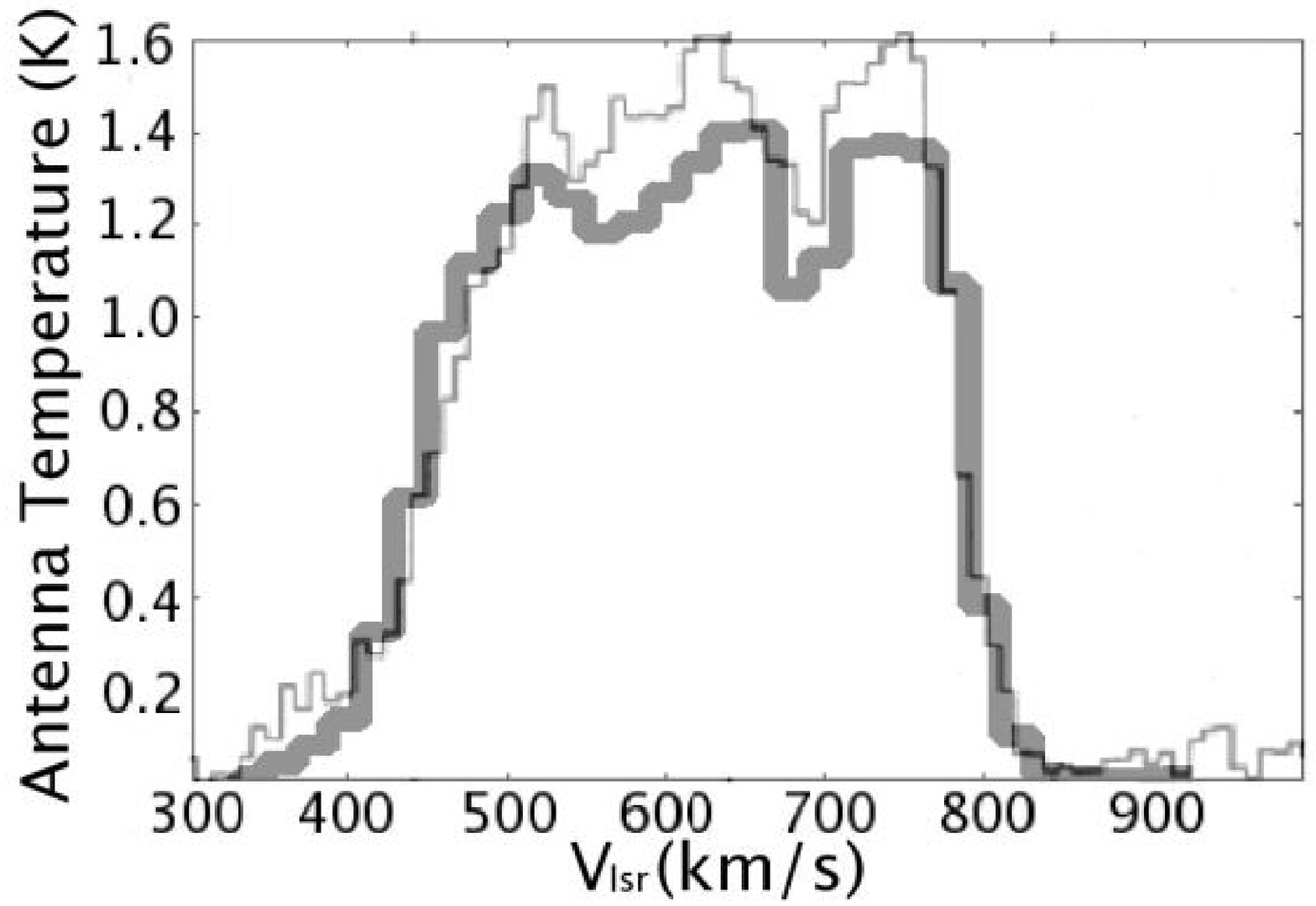}
\end{center}
\caption\small{Line profile in $^{12}$CO(2-1) measured with the SEST \citep[thin line; from][]{hen94} and that measured here with the SMA (thick line).  The latter corresponds to the spatially-integrated line profile corrected for the primary beam of the SMA and convolved to the primary beam of SEST, and using a conversion factor of 0.54 to convert from SMA main beam temperature to SEST antenna temperature \citep{hen94}.  The root-mean-square (rms) noise level of the profiles are 49.6~mK for SEST and 8~mK for the SMA.  The integrated line intensity measured with the SMA is $\sim$$90\%$ (not taking into account any errors in the absolute flux calibration) of that measured with the SEST, indicating that we have recovered the bulk if not all of the emission present in the same region.}
\label{12co-spec-overlay}
\end{figure}

The $^{12}$CO(2-1) emission towards the center of NGC~4945 has been observed a number of times with the SEST \citep{dah93, hen94, cur01, wan04}.  These observations often give different line profiles and intensities, caused most likely by inaccuracies in telescope pointing.  Only the $^{12}$CO(2-1) line profiles measured by \citet{dah93} and \citet{hen94} appear similar, and which also agree with that measured here.  In Figure~\ref{12co-spec-overlay}, we plot our spatially-integrated $^{12}$CO(2-1) line profile corrected for the primary beam of the SMA and convolved to the primary beam of SEST, together with the line profile measured by \citet{hen94}.  The integrated line intensity (in main beam brightness temperature) we measure of $925 \pm 13 {\rm \ K \ km \ s^{-1}}$ is $\sim$$90\%$ of that measured by \citet{hen94} of $1050 \pm 6 {\rm \ K \ km \ s^{-1}}$ \citep[as quoted in][]{wan04}.  The quoted uncertainties do not take into account any errors in flux calibration, and so we have likely recovered the bulk if not all of the $^{12}$CO(2-1) emission present in the same region.

The $^{13}$CO(2-1) and C$^{18}$O(2-1) emissions toward the center of NGC~4945 also have been observed a number of times with the SEST \citep{hen94,cur01,wan04}.  The $^{13}$CO(2-1) line profile we measure is similar in shape to that measured in $^{12}$CO(2-1) as shown in Figure~\ref{12co-spec-overlay}, and also similar to the $^{13}$CO(2-1) line profiles measured by \citet{hen94} and \citet{wan04} but not \citet{cur01}.  The integrated line intensity we measure of $82 \pm 3 {\rm \ K \ km \ s^{-1}}$ is similar to that measured by \citet{wan04} of $81.2 \pm 0.7 {\rm \ K \ km \ s^{-1}}$ \citep[which has a higher S/N ratio than that measured by][]{hen94}.

The C$^{18}$O(2-1) line profile we measure is again similar in shape to that measured in $^{12}$CO(2-1), but is different to those measured by either \citet{cur01} or \citet{wan04}.  The line profile we measure is closest in shape to that measured by \citet{hen94}.  The integrated line intensity we measure of $31 \pm 3 {\rm \ K \ km \ s^{-1}}$ also is similar to that measured by \citet{hen94} of $32 \pm 2 {\rm \ K \ km \ s^{-1}}$.   In Table~1, we summarize the integrated intensities that we measure for all three lines together with the values measured by \citet{hen94}, \citet{cur01}, and \citet{wan04} with the SEST.

\subsubsection{Molecular gas mass from CO lines}
\label{co-mass}

To estimate the mass of molecular gas traced in $^{12}$CO(2-1), we first use the conversion factor between the brightness temperature of the $^{12}$CO(1-0) line and molecular hydrogen column density of $1 \times 10^{20} {\rm \ (K \ km \ s^{-1})^{-1} \ cm^{-2}}$ as has been proposed to be appropriate in the central regions of galaxies \citep[e.g., see][]{pag01}.  Table~\ref{tab:sum-flux} lists the brightness temperature of the $^{12}$CO(2-1) line if its emission as mapped with the SMA had been observed by SEST (details in $\S\ref{subsec:singledish}$).  In Table~\ref{tab:flux-mass}, we list the actual measured brightness of the three observed lines.  Assuming a line ratio (in main beam brightness temperature) of $^{12}$CO(2-1) to $^{12}$CO(1-0) of $1.2\pm0.1$ \citep[from][after correcting for the different beam sizes]{dah93}, we derive a molecular hydrogen column density of $5.9 \pm 0.1\times 10^{22}$~cm$^{-2}$.  Given the measured source size as described in $\S\ref{spatial-kinetic-structure}$ and listed also in Table~\ref{tab:flux-mass}, the corresponding mass in molecular hydrogen gas is therefore $(1.63 \pm 0.03) \times 10^8$~M$_\odot$.

To compute the molecular gas mass from the $^{13}$CO(2-1) and C$^{18}$O(2-1) lines, we assume that this gas is in local thermal equilibrium (LTE).  We adopt the abundance ratio [$^{12}$CO]/[$^{13}$CO]$= 50$ and [$^{12}$CO]/[C$^{18}$O]$= 200$ as estimated by \citet{cur01} and \citet{wan04}.  We assume an excitation temperature of $\sim$30~K based on the peak brightness temperature measured for the $^{12}$CO(2-1) line (i.e., 16.5 Jy/Beam), and that the $^{13}$CO(2-1) and C$^{18}$O(2-1) lines are optically thin.  In this way, we derive a column density from the $^{13}$CO(2-1) line of $(6.7 \pm 0.3) \times 10^{22} {\rm \ cm^{-2}}$, and column density from the C$^{18}$O(2-1) line of $(1.65 \pm 0.17) \times 10^{23} {\rm \ cm^{-2}}$.  Their corresponding molecular hydrogen gas masses are $(1.03 \pm 0.05) \times 10^8$~M$_\odot$\ as traced in $^{13}$CO(2-1) and $(1.21 \pm 0.1) \times 10^8$~M$_\odot$ as traced in C$^{18}$O(2-1).  The estimated mass in molecular gas from all three lines agrees with each other, and are also roughly comparable with the value estimated from the dust continuum emission of $\sim$$4.7 \times 10^8$~M$_\odot$ ($\S\ref{subsec:dustmass}$).

We also compute the approximate mass in molecular hydrogen gas within the region where the disk exhibits circular rotation (i.e., radius $\lesssim 5\farc$).  Table~\ref{tab:flux-mass-small} lists the measured brightness temperature within a Gaussian region of size $10\farc \times 4.7\farc$  and position angle of 44 centered on the disk.  The inferred column density and corresponding mass in molecular hydrogen gas are $\sim 10^{23}{\rm \ cm^{-2}} $ and $\sim 10^7$ \ M$_\odot$\ as listed in Table~\ref{tab:flux-mass-small}.
Follow the same method mentioned above, and the emitting area of 10\farc by 4\farcs7 (assume the inclination of the disk is $62^{\circ}$).  The gas mass derived from $^{12}$CO is $(3.22 \pm 0.06) \times 10^7$~M$_\odot$, and the gas mass derived from $^{13}$CO and C$^{18}$O is $(3.0 \pm 0.1) \times 10^7$~M$_\odot$ and $(3.1 \pm 0.5) \times 10^7$~M$_\odot$, respectively.


\subsubsection{Density and Temperature}
\label{subsec:lineratio}
\begin{figure}[htb]
\hspace{-5cm}
\begin{center}
\includegraphics[width=18cm,angle=0]{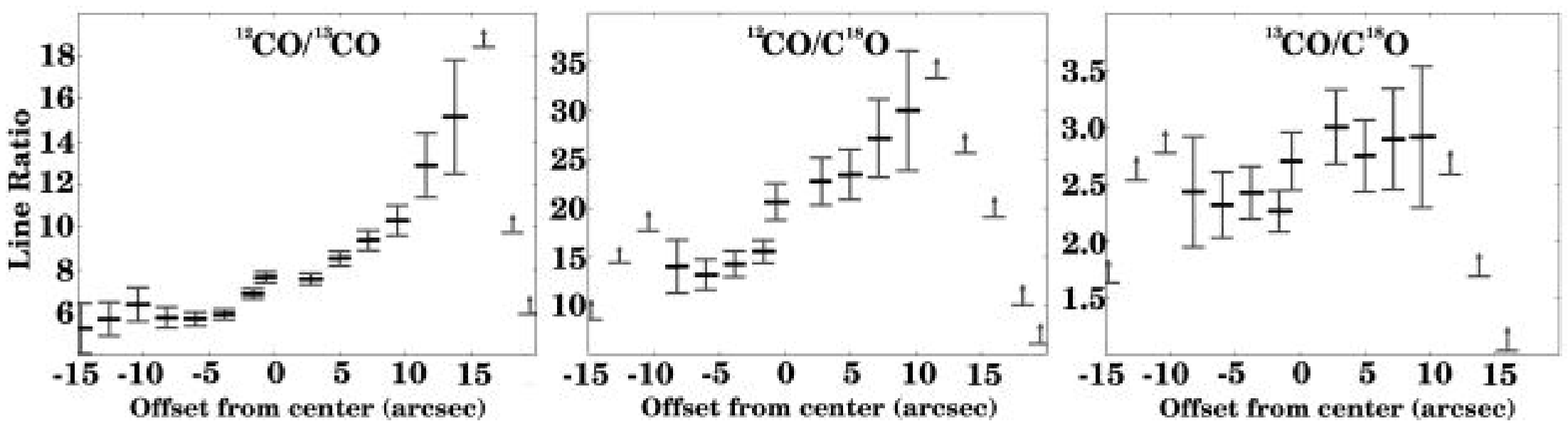}
\end{center}
\caption\small{From left to right: Intensity ratio in $^{12}$CO/$^{13}$CO, $^{12}$CO/C$^{18}$O and $^{13}$CO/C$^{18}$O along the major axis of the disk, measured at intervals of 2\farc\ (half the synthesized beam along this direction).  The error bars indicate $\pm 1\sigma$ uncertainty, and the bars with upward pointing arrows the $2\sigma$ lower limits. The negative offset corresponds to the north-east part of the disk and the positive offset corresponds to the south-west part.}
\label{ratio-major}
\end{figure}

To search for any radial variations in density and temperature of the molecular gas, we computed line ratios (in measured brightness temperature) in $^{12}$CO/$^{13}$CO, $^{12}$CO/C$^{18}$O, and $^{13}$CO/C$^{18}$O along the major axis of the disk.  The results are shown in Figure~\ref{ratio-major}.  Both the $^{12}$CO/$^{13}$CO and $^{12}$CO/C$^{18}$O line ratios increase from the north-eastern (redshifted) side to the south-western (blueshifted) side of the disk, spanning the range 5--15 in $^{12}$CO/$^{13}$CO and 13--30 in $^{12}$CO/C$^{18}$O.  On the other hand, the $^{13}$CO/C$^{18}$O ratio remains roughly constant along the major axis of the disk, occupying a relatively narrow range between 2.2 and 3.0.

\begin{figure}[htb]
\begin{center}
\includegraphics[width=18cm,angle=0]{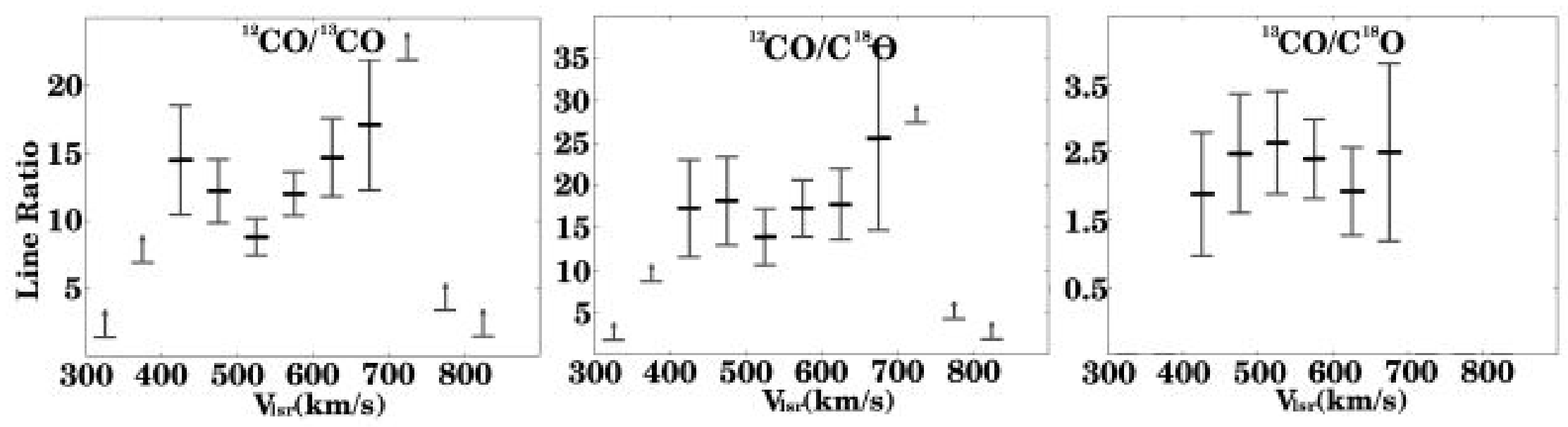}
\end{center}
\caption\small{From left to right: Intensity ratio in $^{12}$CO/$^{13}$CO, $^{12}$CO/C$^{18}$O and $^{13}$CO/C$^{18}$O as a function of velocity as measured from the PV-diagram along the minor axis of the disk (Fig.~\ref{pv-diagrams}, bottom row).  The error bars and the upper limits are the same as in Fig.~\ref{ratio-major}.}
\label{ratio-pvmin}
\end{figure}

To determine the line ratios of the inner spatially-unresolved but kinematically-decoupled component described in $\S\ref{spatial-kinetic-structure}$, we turn to the PV-diagram along the minor axis of the disk as shown in Figure~\ref{ratio-pvmin}.  Notice that the emission from this inner component away from the velocity range 520--650${\rm \ km \ s^{-1}}$, which straddles the systemic velocity, can be well separated from that of the surrounding disk.  In this way, we compute for the inner component line ratios that span the range 12--17 in $^{12}$CO/$^{13}$CO, 17--26 in $^{12}$CO/C$^{18}$O, and 1.2--1.5 in $^{13}$CO/C$^{18}$O.  By comparison, the corresponding line ratios along the minor axis at velocities in the range 520--650${\rm \ km \ s^{-1}}$, where emission from the inner component and surrounding disk cannot be easily separated, are generally lower, spanning the range 8--15 for $^{12}$CO/$^{13}$CO, 14--18 for $^{12}$CO/C$^{18}$O, and 1.2--1.6 for $^{13}$CO/C$^{18}$O.

\begin{figure}[htb]
\begin{center}
\includegraphics[width=10cm,angle=0]{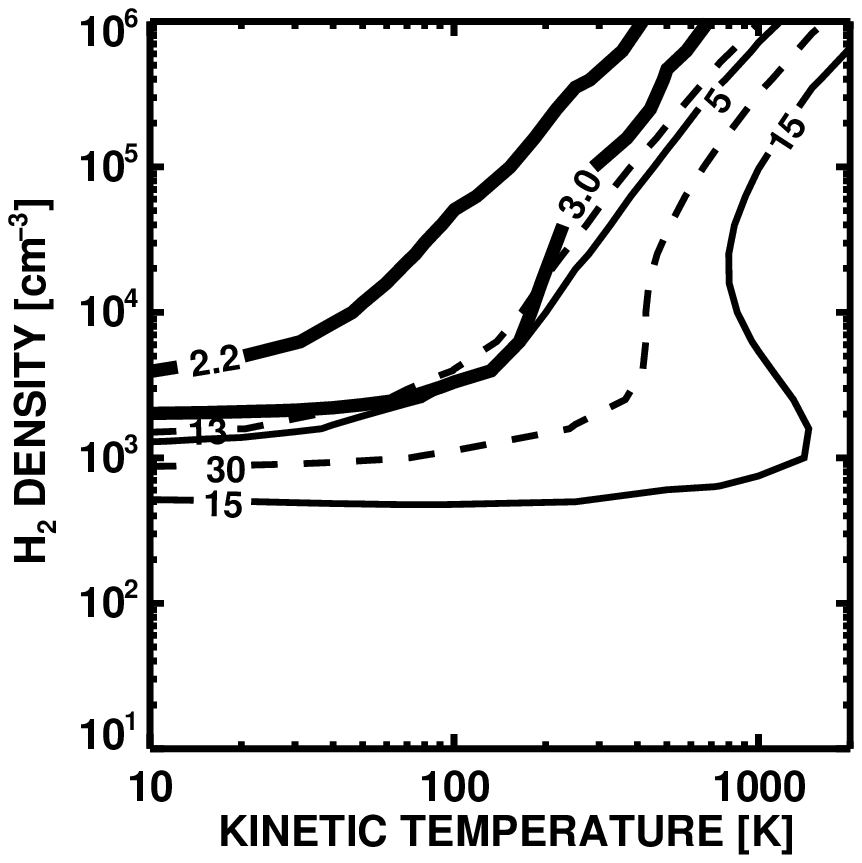}
\end{center}
\caption\small{Results from our LVG calculations for the density and kinetic temperature of the molecular gas in the disk.  The thin lines are solutions derived from the line ratio $^{12}$CO/$^{13}$CO for the values 5 and 15, encompassing the range of values spanned by this ratio.  The dashed lines are solutions derived from the line ratio $^{12}$CO/C$^{18}$O for the values 13 and 30, encompassing the range of values spanned by this ratio.  The thick solid lines are solutions derived from the line ratio $^{13}$CO/C$^{18}$O for the values 2.2 and 3.0, encompassing the range of values spanned by this ratio.  Below a temperature of $\sim$100~K, the line ratio is only weakly dependent on temperature, with smaller line ratios indicating higher densities.  Above $\sim$100~K, for a given line ratio the density increases with temperature, with smaller line ratios continuing to indicate higher densities.}
\label{line_ratios_disk-diagrams}
\end{figure}

We have used the LVG approximation \citep{gol74} to compute the physical conditions of the molecular gas implied by the measured line ratios.  The collision rates for CO in the temperature range $10$--$250$ K were taken from \citet{flo85} and $500$-$2000$ K from \citet{mck82}.  In these calculations, we adopted a relative abundance of $\rm [^{13}CO]/[H_{2}] = 1 \times 10^{-6}$ \citep{sol79}, and isotopic ratios $\rm [^{12}CO]/[^{13}CO] = 50$ and $\rm [^{12}CO]/[C^{18}O] = 200$ \citep{wan04}.  We assume that the emission in all lines are emitted from the same region (i.e., a one-zone model), and a velocity gradient ${\rm dv/dr} \approx 1 {\rm \ km \ s^{-1} \ pc^{-1}}$ given the measured linewidth and diameter for the $^{13}$CO emitting region of $\sim$285~km s$^{-1}$ and $\sim$270~pc respectively.  The results for all three sets of line ratios are shown together in Figure~\ref{line_ratios_disk-diagrams}, and exhibit the following trends: (1) at temperatures $T \lesssim 100 {\rm \ K}$, all three line ratios exhibit only a weak dependence on temperature, with lower ratios corresponding to higher densities; (2) at temperatures $T > 100 {\rm \ K}$, for a given line ratio the density increases with temperature, with smaller line ratios continuing to indicate higher densities; and (3) the line ratios $^{12}$CO/C$^{18}$O and $^{13}$CO/C$^{18}$O tend to indicate a higher density than $^{12}$CO/$^{13}$CO.

The measured line ratios therefore imply a decrease in gas density and column density from the north-eastern (redshifted) side to the south-western (blueshifted) side of the disk.  Our measurements do not place strong constraints on the gas temperature, but we note that the measured $^{12}$CO/$^{13}$CO line ratios are comparable with average values of $\sim$$13\pm5$ found in starburst galaxies \citep{aal95} and $\sim$$13\pm1$ in Seyfert galaxies \citep{pap98}.  If the similar line ratios indicate similar physical conditions, then the preferred solution is for temperatures $T < 100 {\rm \ K}$ and densities ${\rm n(H_2)} \approx 10^3 {\rm \ cm^{-3}}$, typical values of the bulk properties for giant molecular clouds in our Galaxy.  In this regime, the gas density changes by a factor of just a few across the major axis of the disk.  Consistent with this idea, single-dish observations in multiple lines and transitions find a density of $\sim$(3--5)$\times10^{3}$~cm$^{-3}$ and temperature of $\sim$100~K \citep{hen94,cur01,wan04} for the bulk of the centrally unresolved molecular gas.  At radii beyond $\sim$10\farc\ ($\sim$190~pc) at or beyond the outer regions of the disk, where $^{13}$CO and C$^{18}$O are not detectable at our sensitivity limits, the line ratios $^{12}$CO/$^{13}$CO and $^{12}$CO/C$^{18}$O have lower limits that are larger than their measured values at smaller radii.  This indicates that the CO emitting outer region of the disk is dominated by relatively diffuse gas at densities of $\sim$$10^{2} {\rm \ cm^{-3}}$.

The inner kinematically-decoupled component exhibits a $^{13}$CO/C$^{18}$O line ratio that is near unity (1.2--1.5); i.e., the less opaque C$^{18}$O line has nearly the same flux density as the $^{13}$CO line.  The $^{12}$CO/C$^{18}$O line ratio of 14--18 also is relatively low compared with values in the range 40--70 typically found in other galaxies \citep[see][]{sag91,hen93}.  At face value, this suggests that even the $^{13}$CO and C$^{18}$O lines are nearly optically thick in the inner kinematically-decoupled component.  Alternatively, the C$^{18}$O species may be unusually abundant within the inner component.  

We consider the second possibility first.  $^{18}$O enrichment can occur through its production from $^{14}$N by He-burning in high-mass stars ($>8$ M$_{\odot}$), and subsequently dispersed in Wolf-Rayet phase or in type II supernova explosions \citep{sag91,hen93,ama95}.  Such enrichment may have occurred during past circumnuclear starsburt activity in NGC~4945, analogous to that currently seen as the P$\alpha$ ring.  Such low line ratios are seen in at least one other star forming galaxy, NGC~6946 \citep{mei04}.  On the other hand, not all star forming galaxies show such low  line ratios; e.g., both M82 \citep{wei01} and NGC~253 \citep{sak06} exhibit $^{12}$CO/C$^{18}$O or $^{13}$CO/C$^{18}$O line ratios comparable with the mean observed in other galaxies.

\begin{figure}[htb]
\begin{center}
\includegraphics[width=10cm,angle=0]{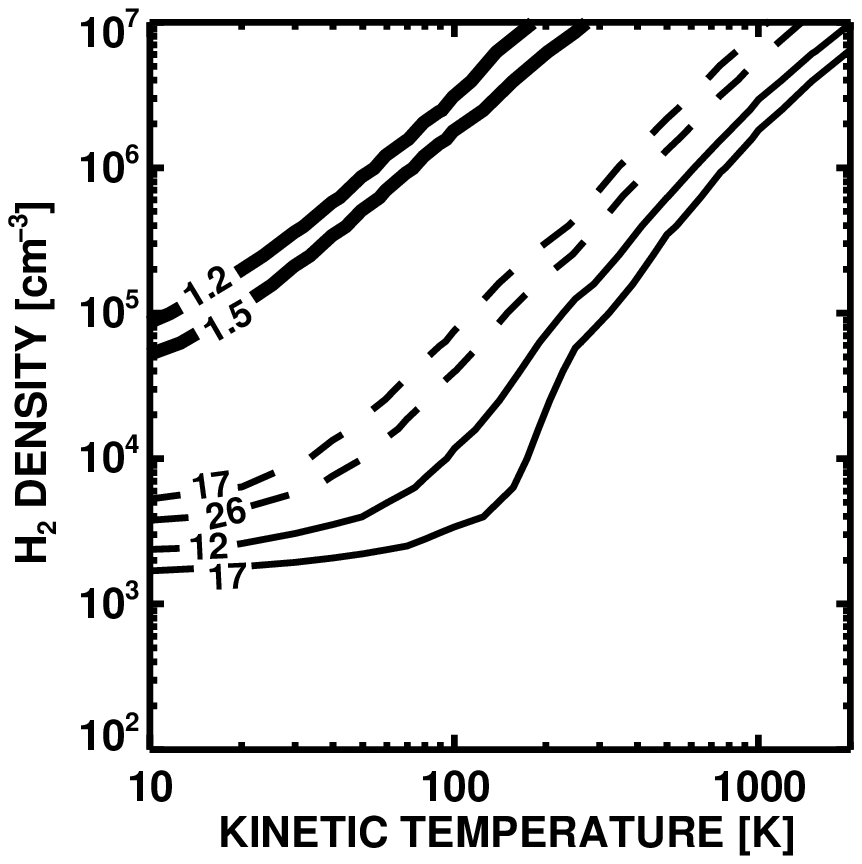}
\end{center}
\caption\small{Results from our LVG calculations for the density and kinetic temperature of the molecular gas in the inner kinematically-decoupled component.  The thin solid lines are solutions derived from the line ratio $^{12}$CO/$^{13}$CO for the values 12 and 17, encompassing the range of values spanned by this ratio.  The dashed lines are solutions derived from the line ratio $^{12}$CO/C$^{18}$O for the values 17 and 26, encompassing the range of values spanned by this ratio.  The thick solid lines are solutions derived from the line ratio $^{13}$CO/C$^{18}$O for the values 1.2 and 1.5, encompassing the range of values spanned by this ratio.  The solutions for $^{13}$CO/C$^{18}$O are quite different from those for $^{12}$CO/$^{13}$CO and $^{12}$CO/C$^{18}$O, indicating that both $^{13}$CO and C$^{18}$O trace a different denser region than $^{12}$CO.}
\label{line_ratios_inner_component-diagrams}
\end{figure}

We turn back to the first possibility, which requires both the $^{13}$CO and C$^{18}$O lines to be optically thick.  To infer the required physical properties of the molecular gas, we use the same LVG approximation and the same assumptions as before for the surrounding disk, except that we now assume an order of magnitude higher velocity gradient as is more appropriate for the inner component.  The results are shown in Figure~\ref{line_ratios_inner_component-diagrams}, which reveals that the physical properties of the gas derived from the $^{12}$CO/$^{13}$CO and $^{12}$CO/C$^{18}$O line ratios are comparable but very different from those derived from the $^{13}$CO/C$^{18}$O line ratio.  Our results imply that the one-zone model is not valid for the inner kinematically-decoupled component, and that the $^{13}$CO(2-1) and C$^{18}$O(2-1) lines trace a different denser region than the $^{12}$CO(2-1) line.  This situation is not uncommon in galaxies, with the $^{12}$CO emission originating from more extended and diffuse gas and the $^{13}$CO or C$^{18}$O emission from more compact and dense gas \citep[e.g.,][]{dow92,wal93,aal95}.  The physical properties of the molecular gas as inferred from the $^{12}$CO/$^{13}$CO and $^{12}$CO/C$^{18}$O line ratios are comparable with those inferred in the surrounding disk, and presumably correspond to the more diffuse part of the inner component.  On the other hand, the measured $^{13}$CO/C$^{18}$O line ratio implies that even at temperatures as low as $T \approx 10$~K, the gas density is $\sim$$5\times10^4$--$1 \times 10^5 {\rm \ cm^{-3}}$.  This is 1--2 orders of magnitude higher than the gas density in the surrounding disk, and presumably corresponds to the denser part of the inner component.  The inferred gas density of the denser part increases towards higher temperatures, with densities about an order of magnitude higher still at $T \approx 100$~K.

\section{DISCUSSION}\label{sec:discussion}

\subsection{Dust Heating}\label{sec:far-ir}
NGC~4945 is one of the three brightest IRAS point sources beyond the Magellanic clouds.  It has a far-IR luminosity of $\sim$$2 \times 10^{10} {\rm \ L_{\odot}}$ \citep{bro88}, which is comparable with that radiated at all other wavelengths combined.  Nearly all (at least 80$\%$) of the far-IR emission arises from a central region no larger than $12\farc \times 9\farc$ ($230 {\ \rm pc} \times 170 {\ \rm pc}$), which is comparable in size to the central continuum source that we detected at 1.3~mm.  The far-IR emission from this central source is attributed to dust at a temperature of $\sim$40~K.

As pointed out by \citet{mar00}, it is not clear whether the central dust-emitting region in NGC~4945 is heated by the AGN or circumnuclear starburst.  This situation reflects the general difficulty in deducing the nature of the source that heats dust in the nuclear region of active galaxies.  In the case of NGC~4945, our observation reveals that the central dust that emits at 1.3~mm (size $\sim$190~pc along the major axis) spans the entire observable extent and somewhat beyond the circumnuclear starburst ring.  This spatially-extended dust is unlikely to be heated predominantly by the AGN as the latter is embedded in obscuring material that prevents UV photons ($13.6 < {\rm h}\nu < 500$~ev) from penetrating beyond a distance of at most 1\farcs5 ($\sim$30~pc) from the center \citep{mar00}.  Even soft X-rays from the AGN can only escape along a X-ray plume (believed to be blown by a nuclear starburst) that emerges north-west of center \citep{sch02}; i.e., the axis of the X-ray plume is orthogonal to the plane of the central molecular gas and dust disk .  Thus, if the dust emitting at 1.3~mm can be used as a proxy for that emitting at far-IR wavelengths, the bulk of this dust is likely heated by the circumnuclear starburst.  Observations at higher angular resolutions are required to search for any dust heated by the central AGN and to study the very inner structure.

\subsection{Central Molecular Concentration}\label{sec:gas-distribution}

\subsubsection{The central disk}
\label{300disk}
Our observation spatially resolves the central molecular gas concentration as traced in CO into an inclined rotating disk.  The radius of this disk, measured over the region where it exhibits rigid-body rotation, is $\sim$5\farc\ ($\sim$95~pc), although the central CO-emitting region extends beyond this radius.  The overall radial size of the emitting region in $^{12}$CO(2-1) is 8\farcs2 (156~pc), which is similar to that inferred from spatially-unresolved single-dish observations by \citet{dah93}.  Their PV-diagram along the major axis of NGC~4945 (their Fig.~3) shows two local intensity peaks at velocities of $\sim$$430 {\rm \ km \ s^{-1}}$ and $\sim$$710 {\rm \ km \ s^{-1}}$ separated by $16\farc \pm 6\farc$, interpreted as the two cross-sections of a highly-inclined ring.  Our PV-diagram (Fig.~\ref{pv-diagrams}) also shows two local intensity peaks near these velocities, specifically at $470$${\rm \ km \ s^{-1}}$\ and $730$${\rm \ km \ s^{-1}}$, separated (as measured from their centroids) by $\sim$13\farc.  These local intensity peaks in the PV-diagram correspond to the location where the rotation curve changes from rigid body to nearly flat, rather than tracing local spatial peaks corresponding to the two cross-sections of an inclined ring.  Our spatially-resolved observations show no evidence for a central hole (or depression) in $^{12}$CO(2-1), nor in $^{13}$CO(2-1) and C$^{18}$O(2-1), thus placing an upper limit of 2\farcs2 (42~pc) on the radius of any such hole.  

The central molecular concentration is therefore a disk or torus (if it has a spatially-unresolved central hole) rather than a ring.  If it has a thickness much smaller than its observed dimensions, then the disk must have an inclination of $62^{\circ}$$\pm2^{\circ}$ to the plane of the sky as measured in $^{13}$CO(2-1), where the disk is least contaminated by surrounding features exhibiting non-circular rotation (see $\S\ref{non-circular-rotation}$ and below).  This is significantly different from the inclination of the large-scale galactic disk of $\sim$$78^{\circ}$.  Observations in atomic hydrogen (HI) gas also indicate a small change in the inclination of the galactic disk with radius (Ott et al. 2001).

As mentioned in $\S\ref{intro}$, \citet{cun05} measured a size for the central molecular concentration in HCN(1-0) of $8\farcs5 \times 4\farcs2$ ($\pm 0\farcs5$) and major axis at ${\rm PA} = 64^{\circ}$.  This is only about half the size that we measured in $^{12}$CO(2-1) of $16\farcs4 \times 10\farcs8$ ($\pm 0\farcs1$).  As in previous observations, \citet{cun05} attribute the observed HNC(1-0) emission to the two cross-sections of an edge-on ring or edge-thickened disk.  Instead, our results suggest that the HNC(1-0) emission originates from the inner region of the disk that we observe here in CO.  The critical density of molecular hydrogen gas for collisional excitation of HNC(1-0) is $\sim$$10^{4-5} {\rm \ cm^{-3}}$, which is about an order of magnitude higher than that for $^{12}$CO(2-1) of $\sim$$10^{3-4} {\rm \ cm^{-3}}$.  The excitation temperatures of these two species, however, are comparable.  The smaller size of the disk in HNC(1-0) compared with that in $^{12}$CO(2-1) therefore suggests a radial decrease in (average) density with radius.  \citet{mau96} reached the same conclusions from a comparison of the line intensities and sizes for the central source in $^{12}$CO(1-0), $^{12}$CO(2-1), and $^{12}$CO(3-2).  In addition, the average gas density on the north-eastern side of the disk is a factor of a few higher than that on the south-western side ($\S\ref{subsec:lineratio}$).

The peaks in HNC(1-0) emission \citep[Fig.~11 of][]{cun05} span approximately the same range of radii as the Pa$\alpha$ starburst ring.  We therefore associate this dense molecular gas with fueling the circumnuclear starburst.  The disk that we observe in CO extends beyond the starburst ring, and therefore traces more diffuse gas in the disk at densities $\sim$$10^3 {\rm \ cm^{-3}}$ ($\S\ref{subsec:lineratio}$).  With an estimated star-formation rate of $\sim$$0.4 {\rm \ M_{\odot} \ yr^{-1}}$ \citep{mo94}, compared with a mass in molecular gas for the disk of $\sim$1--$2 \times 10^8$~M$_\odot$, the circumnuclear starburst must therefore be a transient phenomenon (lasting no longer than $\sim$$10^8$--$10^9$~yrs) if the disk is not replenished from its surroundings.

The dynamical mass of the disk within a radius of 5\farc\ (95~pc), where the disk exhibits circular rotation, is $\sim$$4 \times 10^8$~M$_\odot$.  Using the method described in $\S\ref{co-mass}$, we infer a molecular gas mass within this region of $\sim$$3.1 \times 10^7$~M$_\odot$ (from C$^{18}$O), which is about 13 times lower.  The dynamical mass within a radius of 0\farcs3 (5.7~pc) of the central super massive black hole as inferred from H$_2$O masers is $1.0 \times 10^6$~M$_\odot$\ \citep{gre97}.  The central region of the disk within a radius of 5\farc\ (95~pc) must therefore be dominated in mass by stars or, less likely, by gas not in molecular form.


\subsubsection{A surrounding bar}
\label{non-circular-rotation}
The S-shaped asymmetry in the isovelocity contours of the intensity-weighted $^{12}$CO(2-1) mean-velocity map (Fig.~\ref{mom0+1-maps}) resembles that seen extending to much larger scales in both molecular and atomic hydrogen gas.  \citet{ott01} have imaged the entire disk of NGC~4945 in both $^{12}$CO(2-1) and HI at comparable angular resolutions of $\sim$24\farc\ with the SEST and Australia Telescope Compact Array (ATCA) respectively.  They reported a similar antisymmetric distortion in the isovelocity contours around the central concentration, extending outwards to 100\farc--200\farc\ in HI.  \citet{ott01} attribute these distortions to a (nearly edge-on) bar.  If correct, then this bar must extend inwards (almost) to the central disk, and may be responsible for channeling gas inwards to form this disk.

\subsubsection{The inner kinematically-decoupled component}
\label{inner-component}
The spatially-unresolved but kinematically-decoupled inner component described in $\S\ref{spatial-kinetic-structure}$ spans approximately the same range of velocities as the surrounding disk.  We have conducted the following check to make sure that this feature is not an artifact of the finite angular resolution of our observation.  If within a radius of $\sim$5\farc\ the central concentration can be described by just a disk in rigid-body rotation, then the PV-diagram along its minor axis should show only emission at those velocities within the synthesized beam.  For a radial velocity gradient as measured from the PV-diagrams along the major axis in $^{13}$CO(2-1) and C$^{18}$O(2-1) of $\sim$$27 {\rm \ km \ s^{-1} \ arcsec^{-1}}$, and with a width for the synthesized beam along the major axis of 4\farcs4, the PV-diagram along the minor axis should therefore only show velocities of $520$--$640 {\rm \ km \ s^{-1}}$.  Instead, irrespectiveof the position angle of the cut through the center of the disk, the inner feature spans velocities of $420$--$760 {\rm \ km \ s^{-1}}$.

The gas density of the inner component as inferred in $\S\ref{subsec:lineratio}$ is sufficiently high to excite emission in the $J=1\rightarrow0$ transitions of HCN, HCO$^+$, and HNC as observed by Cunningham \& Whiteoak (2005).  Indeed, their channel map in HNC (which suffers least from line absorption against the central continuum source) shows emission at or near the centroid of the disk even at the largest blueshifted and redshifted velocities of 370 km s$^{-1}$ and 750 km s$^{-1}$.  Similarly, their PV-diagram in HNC along the major axis of the disk shows that, even at disk center, the emission spans a broad range of velocities comparable with the range observed for the inner kinematically-decoupled component.  Because Cunningham \& Whiteoak (2005) do not show a PV-diagram along the minor axis of the disk, however, we cannot be entirely sure that the inner component that we detected in CO also was detected in HNC.

What is the nature of this inner kinematically-decoupled component?  The near-IR vibrational line of molecular hydrogen traces the walls of a conically-shaped cavity with roughly the same lateral size as the Pa$\alpha$ starburst ring, attributed to a superbubble blown by supernova-driven winds \citep{mar00}.  The inner kinematically-decoupled feature may therefore correspond to the cooler component of the molecular outflow.  If extending to the same height as the outflow seen in the near-IR vibrational line of molecular hydrogen ($\sim$3\farcs6), such an outflow would have been barely resolved in our observation (with an angular resolution of 3\farcs3 \ perpendicular to the major axis of the disk).  As mentioned in $\S\ref{subsec:lineratio}$, we infer a density for this inner kinematically-decoupled component that is 1--2 orders of magnitude higher than that of the surrounding disk.  This is contrary to expectations if the inner component corresponds to molecular gas entrained from the disk or surrounding gas.  

A more attractive explanation is that this inner kinematically-decoupled component comprises a separate rotating disk.  This may naturally explain the relatively high density of $\sim$$10^5 {\rm \ cm^{-5}}$ ($\S\ref{subsec:lineratio}$) inferred for the inner component.  The range in velocities exhibited by the inner component is similar to that exhibited by the water masers around the AGN spanning 440--$800 {\rm \ km \ s^{-1}}$ \citep{gre97}, suggesting a connection between the two.  If it has a radius of 1\farcs4 ($\S\ref{spatial-kinetic-structure}$), the dynamical mass of such an inner disk would be $\sim$$2 \times 10^8$~M$_\odot$; if the actual radius is smaller, the corresponding dynamical mass would also be smaller.  By comparison, the dynamical mass inferred from the H$_2$O mega-masers is $1.0 \times 10^6$~M$_\odot$, which is about two orders of magnitude smaller still.  

The column density of intervening neutral gas derived from X-ray absorption towards the AGN is $\sim$$10^{24.7} {\rm \ cm^{-2}}$ \citep{iws93}.  This is about two orders of magnitude higher than the column density of molecular hydrogen gas inferred for the central disk of $\sim$$10^{22.8} {\rm \ cm^{-2}}$ ($\S\ref{co-mass}$).  If the bulk of the absorption originates from the large-scale galactic disk which in HI has a radius of $\sim$11.4~kpc \citep{ott01}, then the average density in this disk is required to be $\sim$$140~{\rm cm}^{-3}$.  This is about two orders of magnitude higher than the typical density in the interstellar medium of $\sim$$1~{\rm cm}^{-3}$, and so the bulk of the X-ray absorption is unlikely to originate from the galactic disk.  Instead, this X-ray absorption may originate from the inner kinematically-decoupled component.  As mentioned in $\S\ref{subsec:lineratio}$, the denser part of inner component has densities of at least $\sim$$5\times10^4$--$1 \times 10^5 {\rm \ cm^{-3}}$.  To produce the required column density in X-ray absorption, this part of the inner component would then have to span a radius of roughly $\sim$20~pc, and hence have a mass of $\sim$$10^7 {\rm \ M_{\odot}}$.  If at this radius the inner component has a rotational velocity of $85 {\rm \ km \ s^{-1}}$, corresponding to half the full range of velocities measured for this component, the enclosed dynamical mass within this radius would then be $\sim$$10^7 {\rm \ M_{\odot}}$, comparable to its estimated mass in molecular gas.

The required properties of the denser part of the inner component necessary to produce the observed X-ray absorption are comparable with the size and rotational velocity (and hence also enclosed dynamical mass) of a highly inclined rotating disk in radio recombination lines imaged by \citet{roy05}.  This disk has its major axis aligned with the large-scale galactic disk, and an ionized gas mass of $10^5$--$10^6 {\rm \ M_{\odot}}$.  If the denser part of the inner component corresponds to the same disk imaged in radio recombination lines, then this disk must be composed primarily of molecular gas.  This dense inner component may therefore be the hypothesized circumnuclear molecular torus invoked by AGN unification models.  In the case of NGC~4945, which harbours a Seyfert~2 nucleus, the circumnuclear molecular torus is required by AGN unification models to be viewed at a large inclination to its rotation axis, as appears to be the case for the disk imaged in radio recombination lines.

\section{Summary and Conclusions}\label{sec:summary}
Previous single-dish maps in the $J=1\rightarrow0$, $J=2\rightarrow1$, and $J=3\rightarrow2$ transitions of $^{12}$CO infer the presence of a spatially-unresolved molecular ring at the center of NGC~4945.  The inferred radius of this ring ranges from $\sim$5\farc\ at the highest transition to $\sim$15\farc\ at the lowest transition.  We have made an interferometric map that properly resolves for the first time the molecular gas as traced in CO at the center of NGC~4945.  Our images in the $J=2-1$ transition of $^{12}$CO, $^{13}$CO and C$^{18}$O and continuum at 1.3~mm show:

\begin{itemize}

\item A disk in all three lines that exhibits rigid-body rotation within a radius of $\sim$5\farc\ ($\sim$100~pc).  Beyond this radius the rotation curve flattens, and the isovelocity contours exhibit an S-shaped distortion that can be traced in $^{12}$CO(2-1) to a radius of 8\farcs2 (156~pc).  There is no central hole in the disk with a radius larger than 2\farcs2  (42~pc).

\item The entire disk has a mass in molecular gas of $\sim$(1--2) $\times 10^8$~M$_\odot$.  The dynamical mass of the disk within a radius of $\sim$5\farc\ ($\sim$100~pc), where it exhibits circular rotation, is $\sim$$4 \times 10^8$~M$_\odot$.  The mass of molecular gas within this radius is $\sim$$3 \times 10^7$~M$_\odot$, and the mass of the central supermassive black hole may not be larger than $\sim$$1 \times 10^6$~M$_\odot$ \citep{gre97}.

\item Based on the measured line ratios, the bulk of this disk likely has densities $\sim$$10^3 {\rm \ cm^{-3}}$ and temperatures $< 100 {\rm \ K}$, in agreement with values inferred from single-dish observations in multiple molecular species.

\item The disk extends beyond the circumnuclear starburst ring seen in Pa$\alpha$.  The major axis of the disk is aligned with that of the starburst ring and also large-scale galactic disk, all of which have $\rm PA \approx 45^{\circ}$.  The disk is inclined by $62^{\circ} \pm 2^{\circ}$ from the plane of the sky, significantly smaller than the inclination of the larger-scale galactic disk of $\sim$$78^{\circ}$.

\item A newly discovered spatially-unresolved and kinematically-decoupled component at the center of the disk in all three lines.  This component spans velocities of 420--$760 {\rm \ km \ s^{-1}}$, which is approximately the same as the range of rotational velocities exhibited by the surrounding disk, a smaller disk imaged in radio recombination lines, and an even smaller disk in water megamasers.  With an upper limit in projected radius as small as 1\farcs4 (27~pc), this component would not have been recognized in previous single-dish maps.  

\item The bulk of the molecular gas in the inner kinematically-decoupled component resides in two very different densities.  The denser portion has a density that is at least one to two orders of magnitude higher than that in the bulk of the surrounding disk.  The more diffuse portion has a density that is comparable to that in the bulk of the surrounding disk.

\item A nuclear continuum source dominated by dust with a radius of 4\farcs9 (93~pc), about half that measured in $^{12}$CO(2-1).  This source extends beyond the bright inner rim of the circumnuclear starburst ring to, and along some directions beyond, its detectable outer radius. 

\end{itemize}

\noindent By comparison, an interferometric map in HNC(1-0), tracing molecular gas at approximately an order of magnitude higher density than the abovementioned CO(2-1) lines, reveals an inclined rotating-disk-like feature with a radius of 4\farc\ (76~pc), interpreted as a ring with an inner radius of 3\farc\ (57~pc) \citep{cun05}.

We interpret our results in the following manner:

\begin{itemize}

\item The primary heating agent of the dust is star formation rather than AGN activity.  \citet{mar00} infer that UV and soft-X-ray photons cannot penetrate beyond a radius of $\sim$1\farcs5 ($\sim$30~pc) from the AGN, whereas the continuum source has a radius about four times larger that is comparable with the outermost extension of the starburst ring. 

\item The S-shaped distortion in the isovelocity contours is likely caused by a bar seen nearly edge-on.  A similar distortion but on larger spatial scales has been seen also in single-dish $^{12}$CO(2-1) maps and interferometric HI maps of the entire galaxy \citep{ott01}.  In HI, this distortion can be traced out to a radius of $\sim$200\farc\ (380~pc), and is attributed to a nearly edge-on bar.  The S-shaped distortion seen in our CO(2-1) map suggests that this bar extends inwards to within $\sim$5\farc\ of the center of the galaxy, and may have been responsible for channeling gas inwards to create or replenish the central molecular disk.

\item The molecular gas mapped in HNC(1-0) traces the inner denser region of the same disk that we map in CO.  The radial extent of the HNC(1-0) gas is approximately the same as that of the circumnuclear starburst ring, and so this relatively dense gas is likely responsible for fueling the circumnuclear starburst.  In the absence of any replenishment, at the present star-formation rate all the available molecular gas in the disk would be consumed in $10^8$--$10^9$~yrs.

\item The denser part of the inner kinematically-decoupled component is a promising candidate for the circumnuclear molecular torus invoked by AGN unification models.  If it is responsible for producing the bulk of the X-ray absorption seen towards the AGN, then this component would have a size comparable with a highly inclined disk imaged in radio recombination lines that has a radius of $\sim$20~pc.

\end{itemize}

Given the measurements now at hand, it would be instructive to construct a dynamical model of NGC~4945 to determine whether the inferred large-scale bar is responsible for channeling gas inwards to replenish the central molecular disk.  As for the inner kinematically-decoupled component, we have conducted follow-up observations at higher angular-resolutions to clarify the nature of this component, and will report the results in a future paper.

We wish to thank all the SMA personnel in Hawaii, Cambridge, and Taipei for their enthusiastic help during the observations.  We also thank A. Marconi for providing near-infrared images of NGC~4945, and M. Wang for the SEST single-dish spectrum.  J. Lim and S. Matsushita acknowledge support from the National Science Council of Taiwan for conducting this work.  The Submillimeter Array is a joint project between the Smithsonian Astrophysical Observatory and the Academia Sinica Institute of Astronomy and Astrophysics and is funded by the Smithsonian Institution and the Academia Sinica.

{\it Facilities:} \facility{SMA}.

\newpage

\begin{table}[h]
\begin{center}
\begin{tabular}{cccc}
\hline
Telescope \& Reference & $^{12}$CO(2-1)$\int T_{mb} dv$ & $^{13}$CO(2-1)$\int T_{mb} dv$ & C$^{18}$O(2-1)$\int T_{mb} dv$  \\
                       & (${\rm K \ km \ s^{-1}}$)   & (${\rm K \ km \ s^{-1}}$)   & (${\rm K \ km \ s^{-1}}$)     \\
\hline
SMA, our results             &  925$\pm$13  &  82$\pm$3     & 31$\pm$3    \\
SEST, Henkel et al.\ (1994)  & 1050$\pm$6   & 123.21$\pm$3.8 & 32$\pm$2.0   \\
SEST, Current et al.\ (2001) &  740$\pm$40  &  86$\pm$5      & 29$\pm$2     \\
SEST, Wang et al.\ (2004)    &  921$\pm$0.6 &  81.2$\pm$0.7  & 24.5$\pm$0.5 \\
\hline
\end{tabular}
\caption{Integrated line intensities measured with the SMA (smoothed to SEST resolution) and SEST.}
\end{center}
\label{tab:sum-flux}
\end{table}

\newpage 

\begin{table}[h]
\begin{center}
\begin{tabular}{ccccc}
\hline
    & Total Flux  &  $\int T_{mb} dv$  & Column Density  &  Hydrogen Gas  \\
    &  (Jy)  & (K kms$^{-1}$)& ($10^{22}$ cm$^{-2}$) & Mass ($10^8$ M$_{\odot}$) \\
$^{12}$CO & 1271 $\pm$ 20 & 713 $\pm$ 11 & 5.9 $\pm$ 0.1 & 1.63 $\pm$ 0.03  \\
$^{13}$CO & 97 $\pm$ 4 & 111 $\pm$ 5 & 6.7 $\pm$ 0.3 & 1.03 $\pm$ 0.05 \\
C$^{18}$O & 29 $\pm$ 3 & 68 $\pm$ 7 & 16.5 $\pm$ 1.7 & 1.2 $\pm$ 0.1 \\
\hline
\end{tabular}
\caption{Emission intensities and derived hydrogen gas mass for the entire disk}
\end{center}
\label{tab:flux-mass}
\end{table}

\newpage

\begin{table}[h]
\begin{center}
\begin{tabular}{ccccc}
\hline
  & Total Flux & $\int T_{mb} dv$  & Column Density &  Hydrogen Gas  \\
  & (Jy) & (K kms$^{-1}$)& ($10^{23}$ cm$^{-2}$) & Mass ($10^7$ M$_{\odot}$) \\
$^{12}$CO & 251 $\pm$ 5 & 2478 $\pm$ 49 & 2.07 $\pm$ 0.04 & 3.22 $\pm$ 0.06  \\
$^{13}$CO & 29 $\pm$ 1 & 323 $\pm$ 15 & 1.94 $\pm$ 0.09 & 3.0 $\pm$ 0.1 \\
C$^{18}$O & 11 $\pm$ 1 & 128 $\pm$ 15 & 4.8 $\pm$ 0.3 & 3.1 $\pm$ 0.5 \\
\hline
\end{tabular}
\caption{Emission intensities and derived hydrogen gas mass within radius $\leq$ 5\farc region}
\end{center}
\label{tab:flux-mass-small}
\end{table}

\newpage

\end{document}